\documentclass[sigconf]{acmart}
\DeclareUnicodeCharacter{0229}{\c{e}}

\AtBeginDocument{%
  \providecommand\BibTeX{{%
    \normalfont B\kern-0.5em{\scshape i\kern-0.25em b}\kern-0.8em\TeX}}}

\newcommand{\name}{\textit{texTENG}}

\usepackage{graphicx}
\usepackage{caption}
\usepackage{subcaption}
\usepackage{xcolor}
\usepackage{soul}

\usepackage[font=small,skip=2pt]{caption}

\setcopyright{acmlicensed}
\copyrightyear{2024}
\acmYear{2024}
\acmDOI{XXXXXXX.XXXXXXX}

\acmConference[AH '25]{Augmented Humans (AHs) International Conference 2025}{March 16--20,
  2025}{Abu Dhabi, UAE}
\acmISBN{978-1-4503-XXXX-X/18/06}

\begin{document}

\title[texTENG: Wearable Textile-Based TENGs]{texTENG: Fabricating Wearable Textile-Based \\Triboelectric Nanogenerators}

\author{Ritik Batra}
\authornote{Both authors contributed equally to the paper}
\email{rb887@cornell.edu}
\orcid{0000-0002-4476-5107}
\affiliation{%
  \institution{Cornell University}
  \city{Ithaca}
  \country{USA}
}

\author{Narjes Pourjafarian}
\authornotemark[1]
\email{np454@cornell.edu}
\orcid{0000-0001-5298-6797}
\affiliation{%
  \institution{Cornell University}
  \city{Ithaca}
  \country{USA}
}

\author{Samantha Chang}
\email{sjc333@cornell.edu}
\affiliation{%
  \institution{Cornell University}
  \city{Ithaca}
  \country{USA}
}

\author{Margaret Tsai}
\email{mpt43@cornell.edu}
\affiliation{%
  \institution{Cornell University}
  \city{Ithaca}
  \country{USA}
}

\author{Jacob Revelo}
\email{jar595@cornell.edu}
\affiliation{%
  \institution{Cornell University}
  \city{Ithaca}
  \country{USA}
}

\author{Cindy Hsin-Liu Kao}
\email{cindykao@cornell.edu}
\orcid{0000-0003-1316-4170}
\affiliation{%
  \institution{Cornell University}
  \city{Ithaca}
  \country{USA}
}

\renewcommand{\shortauthors}{Batra and Pourjafarian, et al.}

\begin{abstract}
  Recently, there has been a surge of interest in sustainable energy sources, particularly for wearable computing.
Triboelectric nanogenerators (TENGs) have shown promise in converting human motion into electric power.
Textile-based TENGs, valued for their flexibility and breathability, offer an ideal form factor for wearables. 
However, uptake in maker communities has been slow due to commercially unavailable materials, complex fabrication processes, and structures incompatible with human motion.
This paper introduces texTENG, a textile-based framework simplifying the fabrication of power harvesting and self-powered sensing applications. By leveraging accessible materials and familiar tools, texTENG bridges the gap between advanced TENG research and wearable applications. We explore a design menu for creating multidimensional TENG structures using braiding, weaving, and knitting. Technical evaluations and example applications highlight the performance and feasibility of these designs, offering DIY-friendly pathways for fabricating textile-based TENGs and promoting sustainable prototyping practices within the HCI and maker communities.

\end{abstract}



\begin{teaserfigure}
  \includegraphics[width=\textwidth]{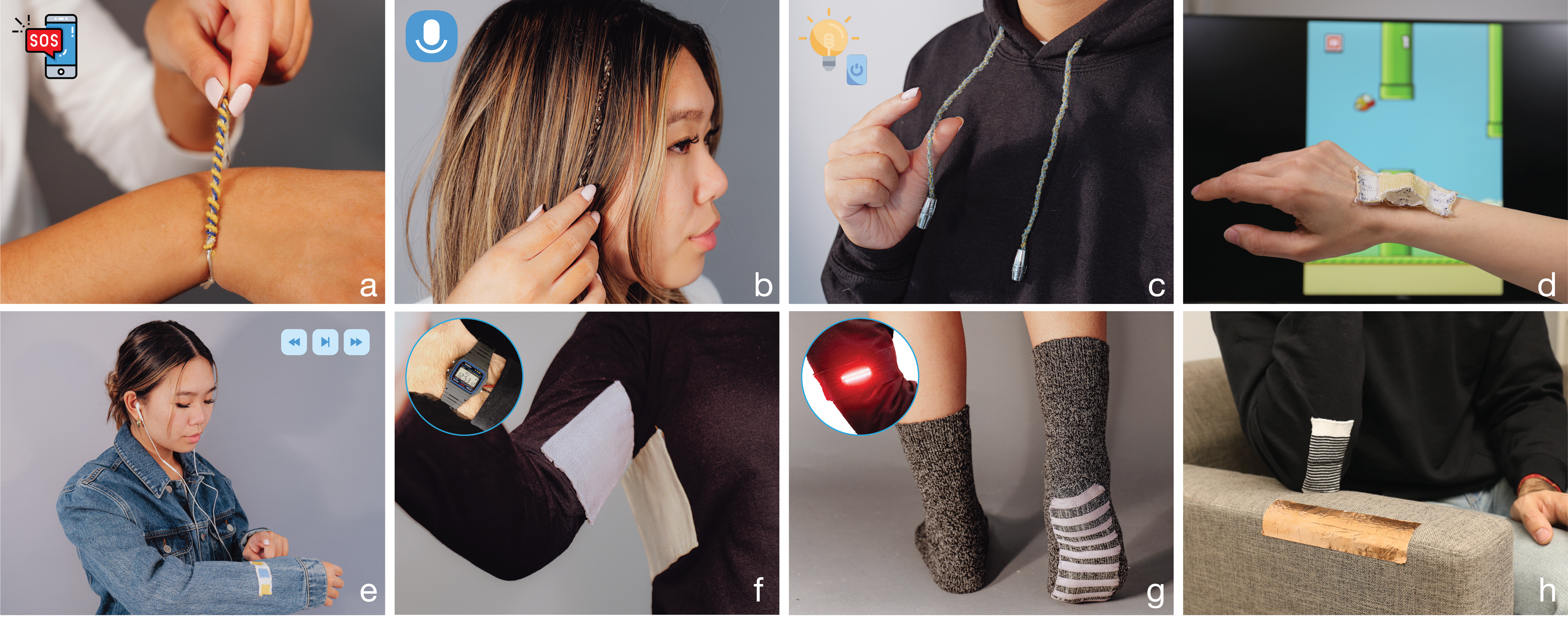}
  \caption{We introduce \name{}, a DIY-friendly framework for fabricating textile-based wearable devices capable of both sensing and power harvesting on the body. Compatible with diverse fabrication approaches and commercially available materials, our menu offers versatility and considerations for the design process. Our application examples demonstrate a) an interactive bracelet for sending emergency messages, b) a smart hair extension for activating voice assistants, c) a touch-sensitive drawstring for light control, d) a game controller utilizing a woven flex sensor, e) self-powered touch sensors for music player control, f) energy storage from smart garments to power a digital watch, g) energy harvesting from smart socks to power running lights, and h) wireless interaction enabled by a knit elbow patch.}
  \label{fig:teaser}
  \Description{}
\end{teaserfigure}

\maketitle

\section{Introduction}
Recent technological advancements and the growing integration of technology into our daily lives have fueled significant interest in wearable computing. As wearable devices become increasingly sophisticated, their higher power consumption presents challenges such as frequent recharging or reliance on bulky batteries. Addressing these issues necessitates more sustainable energy sources, such as triboelectric nanogenerators (TENGs)~\cite{WangTENGHandbook}, which harvest biomechanical energy from human movement and convert it into electric power. This technology also enables self-powered sensing applications, making wearable devices more practical and efficient for everyday use.

Among various mediums, textile-based TENGs offer unique advantages and significant potential as wearable devices. They leverage human motion for energy harvesting while providing the comfort and breathability of textiles.
While current research primarily focuses on enhancing the performance of textile-based TENGs~\cite{ZouOptimizeTENGs} using industrial-grade lab equipment, uptake in maker communities has been slow due to the limited exploration of more user-friendly textile fabrication tools and techniques, commercially available materials, and diverse applications.
Additionally, examining the integration of textile-based TENGs into wearable garments is important in understanding the practical applications of this emerging technology.  

To address these limitations, we present \name{}, a user-friendly framework for fabricating textile-based TENGs for self-powered wearable devices. 
Our work bridges the gap between advanced TENG research and practical applications for the HCI and maker communities. It has the potential to reduce barriers to entry, making technology accessible to a wider audience for experimentation and innovation.
We introduce the HCI and the maker community to the operational principles of textile-based TENGs, their modes and types of operation, and suitable material selections. 
We also present a design menu that synthesizes our investigation of familiar techniques for developing yarn-based (1D), textile structure (2D), and multidimensional TENGs (2.5D) as well as their possible use cases. 

Our exploration of yarn-based TENGs includes examining coating and braiding methods using readily available materials, providing insights into their conversion into wearable accessories. 
Additionally, we discuss suitable textile structures for TENGs using familiar fabrication techniques such as weaving and knitting. We explore mending tools such as DIY darning looms, to integrate these TENGs into existing garments.
Further, we examine multidimensional woven and knitted structures created in a single pipeline without post-processing.
These 2.5D structures, when placed on the body, enable new form factors for sensing and power harvesting that are not feasible with traditional 2D structures, such as flex sensing on joints. 
To our knowledge, we are the first to introduce these 2.5D knitted and woven structures for textile-based TENGs, significantly expanding the design menu for wearable energy harvesting and sensing applications.

We further elaborate on how the harvested energy can be utilized for sensing, actuation, or storage to power other wearable devices. 
Textile-based TENGs afford various applications, including direct integration onto the body surface, incorporation into existing garments, or even serving as the basis for creating textiles and other accessories.
To demonstrate the technical feasibility and practical utility of our design menu in fabricating textile-based TENGs in various fabrication contexts, we introduce eight application examples (Figure~\ref{fig:teaser}). 
These range from 1D accessories, such as interactive bracelets, smart hair extensions, and touch-sensitive drawstrings, to multidimensional self-powered sensors, including touch, flex, and wireless contact sensors.
We demonstrate functional prototypes that harvest energy from everyday movements to power wearables, such as a digital watch from small woven swatches and armband LEDs from a pair of socks.
Furthermore, to gain a better understanding of our approach, we present the results of technical evaluations, showcasing the performance of various TENG materials, efficacy of energy storage, and precision of touch input detection on textile-based TENG devices.

In summary, this paper makes the following contributions: (1) Introducing \name{}, a DIY-friendly framework for textile-based TENGs, tailored for wearable applications, and exploring the design menu for fabricating and seamlessly integrating 1D, 2D, and 2.5D structures using braiding, weaving, and knitting techniques. (2) Presenting a set of example applications compatible with bodily motions that highlight design opportunities for input and energy-harvesting wearable devices. (3) Providing technical evaluations that offer a deeper understanding of the performance and capabilities of the fabricated devices.

We believe that these contributions will advance the understanding of textile-based TENGs and inspire innovative developments and sustainable prototyping practices within the HCI and maker communities.

\section{Related Work}

\subsection{Wearable Computing: Towards Seamless, Conformable, and Invisible Interfaces}

    Mark Weiser envisioned a world where technologies ``weave themselves into the fabric of everyday life until they are indistinguishable from it''~\cite{weiser2002computer}. Wearable devices have since evolved to enhance human augmentation, using the human body as an interface for seamless interaction. Early wearable devices were often in the form of accessories, such as wristbands \cite{Zhang2015tomo, rekimoto2011gesturewrist}, eyeglasses \cite{GoogleGlass}, and rings \cite{Kienzle14lightring, Ogata12}, as these forms could conveniently house electronics. These devices enabled various interactions, from gesture recognition \cite{JiangWearableGestureControl} and haptic feedback \cite{HuangHapticWearable} to audio \cite{olwalsmartglasses} and visual communication \cite{Harrison-OmniTouch}. However, they relied on rigid enclosures and extruded profiles, which were not body-conformable.

    Researchers have since explored alternative wearable interfaces made from softer and more flexible materials, such as silicone \cite{kao2016duoskin, weigel2015iskin, kao2018skinmorph, kao2018skinwire, nittala2020physioskin, lo2016skintillates, Weigel2016skinmarks, nittala2018multitouchskin}, ink~\cite{ChoiBodyPrinter, PourjafarianBodyPrinter}, flexible electronics~\cite{KaoNailO}, and textiles~\cite{ZhuBioWeave, ZhuETextileProduction}.      
    Textiles provide one of the most fundamental interfaces for wearable computing due to their ubiquity on and around the human body~\cite{SinghaWearableSmartTextiles}. 
    Researchers in HCI have explored textile-based interfaces for input sensing \cite{parzer2018resi,parzer2017smartsleeve,leong2016procover, zeagler2012jogwheel,karrer2011pinstripe, IOBraid,schneegass2015simpleskin,chan2017data} and output actuation \cite{williams2015swarm, granberry2019sma,kuusk2015lace}.  
    Despite the advancements in conformable substrates for textile interfaces, the \emph{power source} remains a major barrier to achieving Weiser's vision of \emph{invisible computing} in wearable devices. Current power sources, predominantly Lithium-Polymer (Li-Po) batteries, are bulky, limited in charge capacity, and made from toxic materials. While alternative, conformable, energy harvesting approaches have been examined in the material sciences~\cite{acssuschemeng,he2024flexible}, 
    they have yet to reach the broader HCI and maker community to enable widespread exploration into daily applications. 
    
    Our research aims to overcome this barrier by facilitating the user-friendly fabrication of sustainable and body-conformable power sources and self-powered sensors using textile-based TENGs. Leveraging the inherent flexibility and breathability of textiles, our approach enables the integration of TENGs into garments and accessories.
    While we acknowledge that the harvested energy from wearable TENGs is currently insufficient to replace batteries for most applications, we believe introducing user-friendly fabrication approaches to the HCI and maker community will pave the way for enhancing the performance of textile-based TENGs in the near future. We aim to contribute to the design and fabrication of sustainable HCI~\cite{disalvo2010sustinablehci}, through the making of eco-friendly power solutions for wearable computing.

\subsection{Smart Textile Fabrication Techniques in HCI}
    Various techniques exist for fabricating smart textiles at the yarn level (1D) and in multi-dimensional (2D and 2.5D) structures. Yarn-level research (1D) in HCI has advanced sensing~\cite{ZhuBioWeave,olwal2020textile,parzer2018resi, ZhangCarbonNanotube, IOBraid,ShahmiriSerpentine} and actuation~\cite{forman2019modifiber,kilic2021omnifiber, FibeRobo_Forman_UIST23,DevendorfPlyingSmartTextiles,DierkHairIO} capabilities through techniques such as braiding~\cite{IOBraid, DierkHairIO}, coating~\cite{ZhuBioWeave}, and twisting~\cite{DevendorfPlyingSmartTextiles}. The yarn-based form factor is versatile and can be integrated into various textile surfaces.
    To create a 2D smart textile, the fabrication processes involve \emph{surface-level} and \emph{structural-level} integration of interactive elements into the substrate. \emph{Surface-level integration} includes stitching \cite{perner2010handcrafting,buechley2006lilypad}, embroidery \cite{orth2018ubicomp, zeagler2012jogwheel,hamdan2018chi,he2020plotting,goveia2020crafting,goudswaard2020fabriclick, Jiang2024embrogami, aigner2020embroidered}, machine sewing \cite{nabil2019seamless}, felting \cite{berzowska2005kukkia,hudson2014teddybear}, silk-screening \cite{kim2010screenprinttextile, kuznetsov2018ScreenprintTEI}, and inkjet printing \cite{yoshioka2006inkjetprint}. These methods can rapidly augment a textile but only alter it extrinsically.
    
    Weaving~\cite{bonderover2004woveninverter,dhawan2004wovencircuits1,ivan2016chi,DevendorfPlyingSmartTextiles,friske2019adacad,sun2020wovenskin,WuLoomPedalsTEI} and knitting~\cite{wijesiriwardana2004knitcircuit, OuSensorKnit, granberry2019sma, albaugh2019softactuatedobjects} enable \emph{structural-level integration} of interactive elements. Weaving intersects yarns perpendicularly on a loom, making it ideal for integrated circuits~\cite{bonderover2004woveninverter, dhawan2004wovencircuits1, dhawan2004wovencircuits2}, touch surfaces \cite{ivan2016chi,parzer2017smartsleeve,zysset2010weaving,wu2020zebrasense}, morphing interfaces \cite{tomico2013crafting}, and textile displays \cite{devendorf2016chi,DevendorfPlyingSmartTextiles,berzowska2010karma}. Darning looms, a hand loom traditionally used for mending, enable a more sustainable method of incorporating electronic components into existing garments~\cite{leeEDarningTEI}. Knitting, which involves interlocking loops of yarn, creates elastic fabrics suitable for sensing touch~\cite{VallettCapacitiveTouchFabric}, force~\cite{PointnerKnittedResi}, and stretch~\cite{OuSensorKnit}, as well as for robotic wearables providing haptics~\cite{albaugh2019softactuatedobjects, granberry2019sma, granberry2017activestocking, kim2021knitdermis, kim2024mediknit} and locomotion \cite{kim2022knitskin}.
    Multidimensional textiles can be realized by increasing the number of textile layers~\cite{Devendorf-adacad, Woven-eTextiles-in-HCI_Pouta_DIS22, wu2020automatic}. Weaving with multiple intertwining layers enables interactive capabilities such as sensing~\cite{ZhuBioWeave} and on-skin actuation~\cite{Pin-Sung-PATCH-o}. Knitting can also achieve multi-layering through structures such as tubular~\cite{wicaksono3Dknits2022}, pleats~\cite{Greinke-Folded}, or spacer fabrics~\cite{albaugh2021engineering}.
    
    This paper builds upon existing techniques by introducing comprehensive approaches for fabricating textile-based TENGs using familiar crafting techniques such as braiding, weaving, and knitting. We offer a simplified and user-friendly pipeline for creating multi-dimensional textile structures. Through this, we aim to broaden access to the fabrication of textile-based energy harvesting and sensing devices. This enables makers and HCI researchers to produce self-powered systems without needing advanced material science expertise or sophisticated lab equipment.

\subsection{Triboelectric Nanogenerators in HCI}

    The constant motion of humans makes mechanical energy a promising source for energy harvesting in wearable devices. Traditional methods such as piezoelectric~\cite{HossainPiezoelectricTextile} and electromagnetic~\cite{LeeElectromagneticWearable} harvesting often suffer from low efficiency and require inaccessible materials, making them unsuitable for wearable applications. TENGs have emerged as a superior alternative due to their ability to generate power from a diverse array of materials and adaptable form factors~\cite{WangTENGHandbook}.
    Recently, the HCI community has shown interest in TENGs for their sustainable energy-harvesting potential from everyday body motion. Initial research has demonstrated how TENGs can facilitate touch sensing~\cite{WuIWood, TriboTribe, Shi-WooDowel} and audio detection~\cite{AroraSATURN, AroraZeusss}. These applications necessitate interfaces that are soft~\cite{Skinergy_UIST23} and deformable~\cite{ShahmiriSerpentine, ChenSPIN} to conform to body geometry and avoid restricting natural movement.
    
    As TENGs are integrated on or around the body, wearability factors such as comfort, stretchability, breathability, and material efficiency must be considered~\cite{gemperle1998designforwearability}. Textile-based TENGs are particularly well-suited to meet these requirements, yet they remain underexplored in HCI field~\cite{GowrishankarTENGTextilesWeaving}. There is a clear opportunity to advance the development of textile-based TENGs, which offer the dual benefits of energy harvesting and conformability for wearable technologies.
    
    We synthesize these considerations into a comprehensive design menu that empowers makers and researchers to explore a wide range of materials and techniques tailored to their specific use cases. This practical orientation significantly lowers the barrier to entry for TENG research, enabling broader participation and innovation in the field.
    By democratizing textile-based TENG fabrication techniques, \name{} pushes for the development of sustainable and user-friendly wearable technologies.

\section{Background: Triboelectric Nanogenerators}

\name{} operates based on triboelectric nanogenerators (TENGs), which convert mechanical energy into electricity for sensing and harvesting~\cite{WangTENGHandbook,Zhang-StructuresOfTENG-2019}. 
Below, we distill the principles of TENGs' operations for the HCI and maker community, to enable easier adoption and innovation. Understanding these principles is necessary to better understand the design decisions. 

\subsection{Principles of Operation}

\begin{figure*}[ht]
\includegraphics[width=\linewidth]{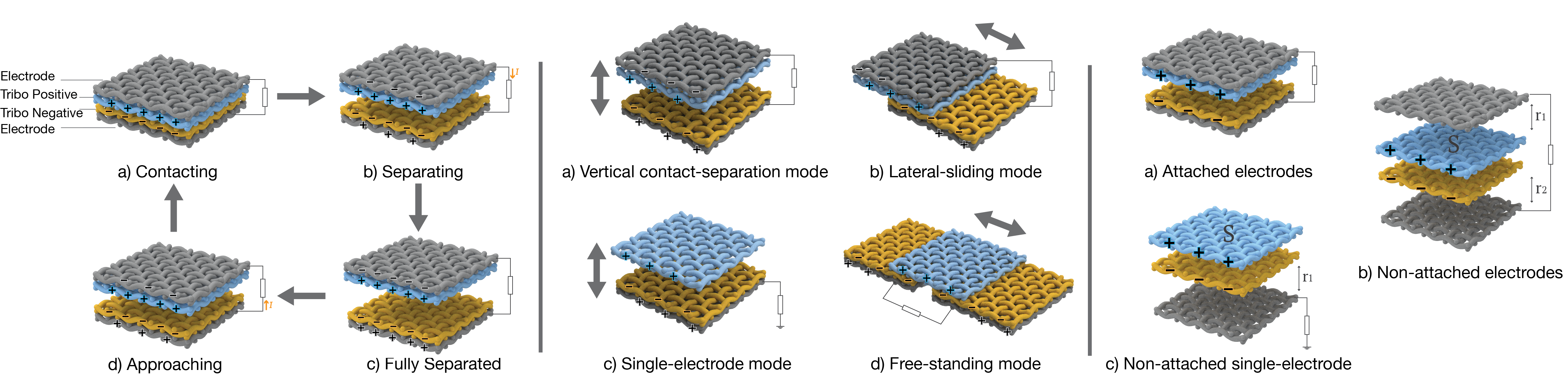}
  \caption{Left) Operational principle of TENGs; Center) Four fundamental operating modes of TENGs; Right) Different types of TENGs based on the bonding between electrodes and tribo layers.}
  \label{fig:background}
  \Description{}
\end{figure*}

TENGs convert mechanical energy into electricity through contact electrification and electrostatic induction between two dielectric layers (positive and negative) fixed to conductive electrodes. When the layers contact, they generate equal and opposite charges (Figure~\ref{fig:background}a-Left). Upon separation, these charges induce a potential difference, causing electron flow (Figure~\ref{fig:background}b-Left). As the gap widens, the electric potential stabilizes, and the charges act as an electrostatic induction source (Figure~\ref{fig:background}c-Left). 
Upon the layers approaching each other again, electrons flow back to compensate for the electrical potential differences (Figure~\ref{fig:background}d-Left).
Repeated contact and separation cause electrons to oscillate between electrodes, generating an alternating current (AC) in the circuit, characterizing TENG operation~\cite{Wang2023-chapter1}.

\subsection{Modes of Operation}        
Various types of TENGs have been designed and fabricated, categorized into four basic modes: vertical contact-separation, lateral sliding, single electrode, and freestanding~\cite{DongFiberFabricTENG}. These diverse modes of TENGs have the potential for versatile HCI applications as portable and wearable power sources. 
    
\subsubsection{Vertical contact-separation (CS) mode (Figure~\ref{fig:background}a-Center)}
This mode is characterized by the presence of two distinct tribo-materials facing each other and separated by a specific distance. Electrodes are attached to each tribo-surface, and the relative motion is perpendicular to the interface. 
    
\subsubsection{Lateral-sliding (LS) mode (Figure~\ref{fig:background}b-Center)}
The lateral-sliding mode operates on the sliding contact between two dielectric surfaces. In a state of perfect alignment, no current flows, as positive charges on one side are entirely offset by negative charges. However, introducing relative displacement through an externally applied force in the direction parallel to the interface generates a potential difference across the two electrodes.
The periodic changes in the effective contact area generate triboelectric power.

\subsubsection{Single-electrode (SE) mode (Figure~\ref{fig:background}c-Center)}
In SE mode, the simplest mode of TENGs for integration, one layer is a reference electrode connected to the ground, and another layer (e.g., human skin) acts as the moving triboelectric layer. As the charged dielectric approaches and moves away from the grounded surface, an induction current is generated and flows back. This mode is advantageous for harvesting energy from moving objects, such as walking and typing, without needing electrodes on both sides. However, it has relatively low output power compared to other modes.

\subsubsection{Free-standing (FS) mode (Figure~\ref{fig:background}d-Center)}  
In FS mode, a dielectric layer moves freely between symmetric electrodes. As it slides, it induces an asymmetric charge distribution, causing electrons to flow between the electrodes to balance the potential. This electron oscillation generates an alternating current (AC).
For \name{}, we explored the first three modes of operation that are more relevant for wearable devices with various application examples (see Section \ref{application}). 

\subsection{Types of Electrodes Placement}

Traditional TENG configurations limit mobility due to direct bonding between tribo surfaces and electrodes (Figure~\ref{fig:background}a-Right). To address this, new TENG designs decouple electrodes and tribo surfaces, enabling near-field electrostatic induction for wireless applications (Figure~\ref{fig:background}b-Right) \cite{VeraAnayaNearFieldTENGs, SHI2017479}. Non-attached single-electrode configurations (Figure~\ref{fig:background}c-Right) allow for remote applications like gesture recognition. 
The performance of these devices depends on the distance between electrodes ($r_1$ and $r_2$) and tribo surfaces ($S$). For \name{}, we explored both conventional and non-attached configurations with various fabrication methods (see Section~\ref{application}).

\section{Design Menu for Fabrication Strategies of Textile-based TENG}

\begin{figure*}[ht]
  \centering
  \includegraphics[width=\linewidth] {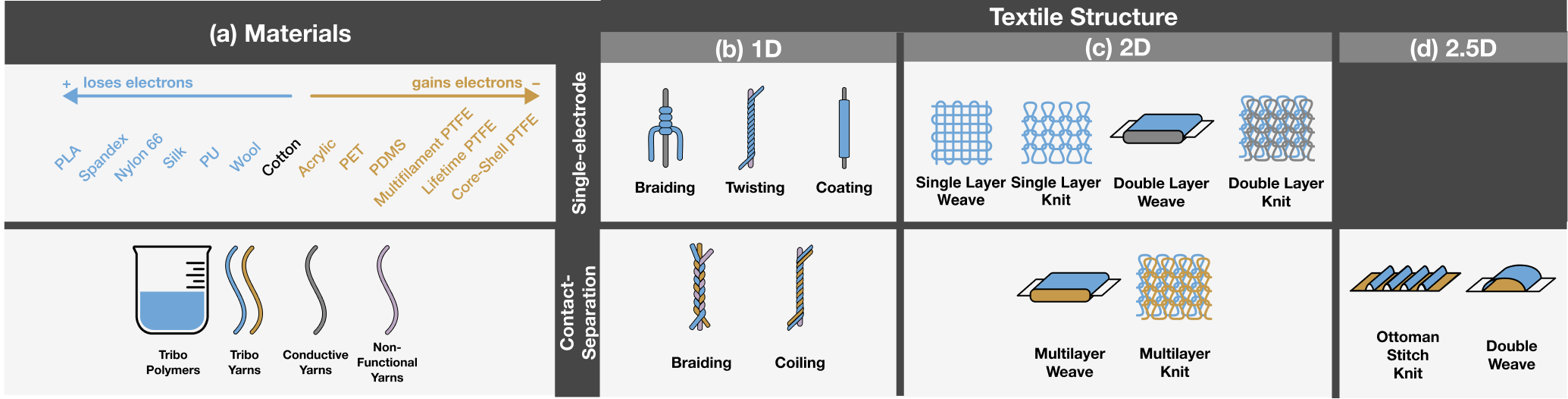}
  \caption{Design Menu of \name{} includes (a) commercially available triboelectric materials and textile structures for fabricating textile-based TENGs, including (b) 1D yarn-level structure, (c) 2D textile structures, and (d) 2.5D textile structures. Materials are ordered based on our experiments and insights from Liu et al.~\cite{LIU2018383}. (Since separating the electrode from the tribo layer does not enhance TENG performance, we opted not to incorporate single-electrode 2.5D structures in our implementation.)
  }
  \label{fig:DesignSpace}
  \Description{}
  \end{figure*}

For the fabrication of textile-based TENGs, we have distilled a design menu (Figure~\ref{fig:DesignSpace}) for user-friendly fabrication. 
The design menu is generated from a thorough literature review~\cite{DongFiberFabricTENG,TianTextileBasedTENGs}, preliminary interviews with three expert fiber artists, and a one-year, iterative hands-on prototyping process involving a team of six researchers with backgrounds spanning HCI, engineering, and textile fabrication. 
Our exploration involved identifying triboelectric materials and user-friendly fabrication approaches suitable for \name{}. 
We offer design recommendations for wearable devices within each component of the proposed design menu.

\subsection{Informing the Design Menu: Insights from Fiber Artists}
To refine our design menu, we conducted preliminary interviews with three expert fiber artists: two weavers (P1 - 16 years and P2 - 9 years of experience) and one machine knitter (P3 - 25 years of experience). Each 30-minute interview, conducted via Zoom, aimed to bridge the gap between advanced TENG research and the needs of textile makers.

The interviews began with discussions about the participants’ craft expertise and familiarity with smart textiles. We introduced the principles of TENGs and then asked the artists to reflect on how this technology could integrate into their practices. Their responses highlighted a critical disconnect between the technical complexity of TENGs and the accessible techniques familiar to craftspeople. As P1 noted, \emph{``Even though I've worked on e-textiles, I still don't really get it. I mean, how the electrical part works.''} a sentiment echoed by the other interviewees.

To address this gap, we invited the artists to share how they would approach fabricating TENGs based on the provided background knowledge. These conversations were transcribed and analyzed using thematic analysis~\cite{braun2012thematic}, revealing four key themes: \textbf{body location}, \textbf{material selection}, \textbf{textile structures}, and \textbf{fabrication techniques}. These themes provided valuable insights into adapting TENG fabrication for broader accessibility.

\noindent \textbf{Body location.}
Participants suggested placing TENGs on highly mobile areas, such as joints (e.g., elbows and knees), due to the constant motion in these regions. As P2 explained, \emph{``I was thinking about...different joints, for example, like on the arm...[and] attaching [TENGs] to the sleeve of an existing garment...because there are a lot of movements around that.''} P2 also expressed interest in capturing energy from more subtle movements, such as leg shaking or tapping.

\noindent \textbf{Material selection.}
Sustainable materials such as silk (P1, P2) and recycled acrylic (P1, P3) were favored due to their practicality and familiarity, aligning with the makers' existing workflows. Additionally, these fibers offer comfort and aesthetic appeal, which are critical for wearable applications. 

\noindent \textbf{Textile structures.}
P3 emphasized the importance of finding the right structure for specific body parts, particularly those that move frequently. For joints, structures that can \emph{``bounce back''} and accommodate flexibility were seen as essential. The makers highlighted the advantages of woven textiles for TENGs, describing them as \emph{``inherently balanced''} and well-suited for durability and multi-layer integration. Conversely, they found the stretchiness of knitted fabrics ideal for areas requiring flexibility, such as elbows or knees.

\noindent \textbf{Fabrication techniques.}
Interviewees identified weaving as particularly compatible with TENGs due to its stability and ability to support multiple layers. Knitting, on the other hand, was preferred for body areas requiring elasticity. Braiding was seen as a promising technique for creating visually appealing accessories like bracelets or hair extensions.

These insights, combined with our literature review and prototyping experience, informed the development of our design menu. By addressing material selection, body location, textile structure, and fabrication techniques, this menu provides guidance for makers seeking to integrate TENGs into wearable applications. The proposed design menu ensures accessibility and ease of adoption while empowering a broader audience to explore TENG-enabled smart textiles.

\subsection{Triboelectric Materials (Figure~\ref{fig:DesignSpace}a)} 

The choice of triboelectric materials significantly impacts the performance of textile-based TENGs.
Pairing materials with opposite triboelectric polarities enhances the electrical output of TENGs, with those further apart on the triboelectric series~\cite{zou2019quantifying} demonstrating superior charge transfer capabilities.
However, not all materials in the conventional triboelectric series are suitable for textile fabrication.
Therefore, we developed a textile-based triboelectric series tailored for wearable devices and DIY fabrication techniques.
This series was curated through prototyping to evaluate each material's performance, compatibility with textile structures, and fabrication durability.
The resulting series includes a diverse range of yarns and polymers that meet functional and aesthetic requirements for practical applications.

Triboelectric materials come in various forms, such as liquids (e.g., PDMS) and yarns (e.g., silk). Liquid materials can coat other yarns, transforming them into yarn form. Once in yarn form, techniques like braiding, weaving, and knitting can create the desired TENG.
Conductive materials are crucial for TENG design, requiring good electrical conductivity and mechanical properties. Metals, metallic derivatives, and conductive polymers are common choices, each with unique advantages and challenges. Non-functional yarns, typically made of triboelectric-neutral cotton, can be used for aesthetics without impacting performance.

\subsection{Textile Structure: 1D TENG Structures (Figure~\ref{fig:DesignSpace}b)}
\begin{figure*}[ht]
  \centering
  \includegraphics[width=\linewidth]{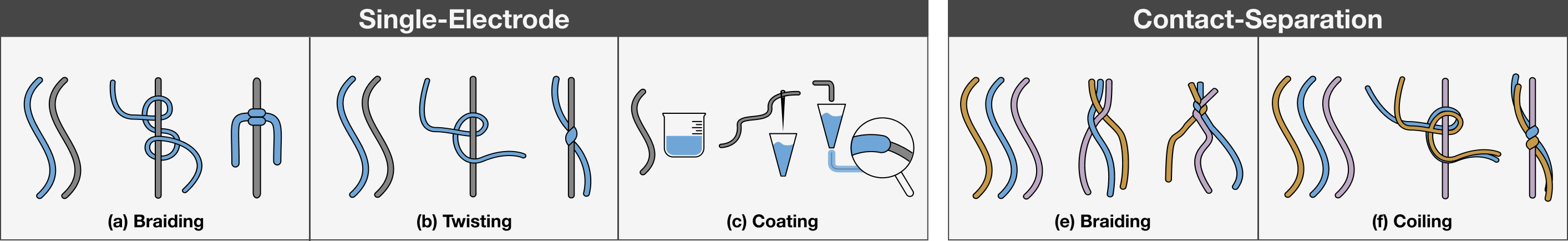}
  \caption{DIY techniques for fabricating yarn-based TENGs (blue and yellow represent positive and negative tribo yarns, respectively, on the triboelectric series, gray represent conductive yarns, and purple represent non-functional yarn).}
  \label{fig:1dyarn}
  \Description{}
\end{figure*}

There are diverse approaches for fabricating and integrating TENG devices into textiles. Yarns provide the fundamental building blocks of textile-based TENGs and can be categorized into two types based on their operational modes: \textit{single-electrode (SE)} and \textit{contact-separation (CS)} configurations.    
In a \textit{SE yarn-based} configuration, a conductive core is sheathed by a triboelectric outer layer, inducing the triboelectric effect while acting as an electrical insulator for the central electrode. Fabrication techniques for \textit{SE yarns} include coating \cite{High-Energy-Asymmetric-Supercapacitor}, braiding \cite{NingFiberTENGRespiratoryMonitoring}, and twisting \cite{Yu2017CoreShellYarnBasedTN} (Figure~\ref{fig:1dyarn}-SE).

\subsubsection{Braiding (SE)}
Our braided yarns consist of a silver-coated conductive yarn as the core, with one polarity of tribo yarns (e.g., PTFE) manually braided around it. The conductive yarn is insulated using macrame techniques such as the half-hitch knot (Figure~\ref{fig:1dyarn}a).    
In addition, we employed a simple DIY braiding tool\footnote{\href{https://www.coolmaker.com/en\_us/toys/778988432426}{https://www.coolmaker.com/}} and fed two spools with tribo yarns, two with insulated conductive yarns, along with two cotton yarns for aesthetic purposes (Figure~\ref{fig:fabMethods}a). The number of feed spools used varies depending on the chosen structure and directly impacts the performance of the braided yarns.
While these techniques are ideal for small-scale fabrication, such as creating accessories, for larger-scale fabrication, commercial braiding machines \cite{NingFiberTENGRespiratoryMonitoring} or DIY tools \cite{FibeRobo_Forman_UIST23} can be utilized.    

\subsubsection{Twisting (SE)}
We fabricated a twisted core-shell structure yarn by tightly twisting the shell (tribo) yarns around the core (conductive) yarn manually (Figure~\ref{fig:1dyarn}b).
For larger-scale production, a twisting machine \cite{Yu2017CoreShellYarnBasedTN} can be used. The flexibility and strength of the braided and twisted yarns depend on factors such as the material, diameter, and number of both the core and shell yarns. 

\subsubsection{Coating (SE)}
We utilized commercially available coated yarn (e.g. core-shell PTFE (Maeden)) and also produced our own coated samples through DIY techniques.
For our DIY coating method, we employed Polydimethylsiloxane (PDMS) (Sylgard 186; Dow Corning (US)). The PDMS and curing agent were mixed at a 10:1 weight ratio to form the coating material. Subsequently, a conductive stainless steel thread (Sparkfun) was coated using a pipette tip, which was filled with the PDMS mixture. Using a needle threader, the conductive thread was carefully passed through the end of the pipette tip. The thread was then slowly pulled through the tip, ensuring complete coating in the PDMS mixture (Figure~\ref{fig:1dyarn}c).
Finally, the coated thread was cured in an oven for 30 minutes at 100 degrees Celsius to solidify the coating.
 
The \textit{CS yarn-based} configuration involves both positive and negative tribo-materials, maximizing energy conversion efficiency, especially in applications limited to a linear form factor. Various fabrication techniques, including coating \cite{cheng2017stretchable}, braiding and coiling \cite{NingFiberTENGRespiratoryMonitoring}, have been developed for \textit{CS yarn-based} TENGs using industrial-grade lab equipment (Figure~\ref{fig:1dyarn}-CS). We aim to make the fabrication user-friendly for maker and HCI communities.
    
\subsubsection{Braiding (CS)}
We created a \textit{CS yarn} by hand-braiding one positive (e.g., PU) and one negative (e.g., PTFE) \textit{SE yarn} (Figure~\ref{fig:1dyarn}e). This method allows customization for enhanced signal strength or aesthetic variation by adjusting the number of strands. Commercial and DIY braiding machines can replicate this structure.

\subsubsection{Coiling (CS)}
To create a more elastic structure, we manually coiled one positive and one negative braided \textit{SE yarns} around an elastic yarn (e.g., Spandex) as shown in Figure~\ref{fig:1dyarn}f. The resulting yarn exhibits excellent stretchability performance. This process can be automated using automatic winding machines~\cite{NingFiberTENGRespiratoryMonitoring}.

\begin{figure*}[ht]
  \centering
  \includegraphics[width=\linewidth]{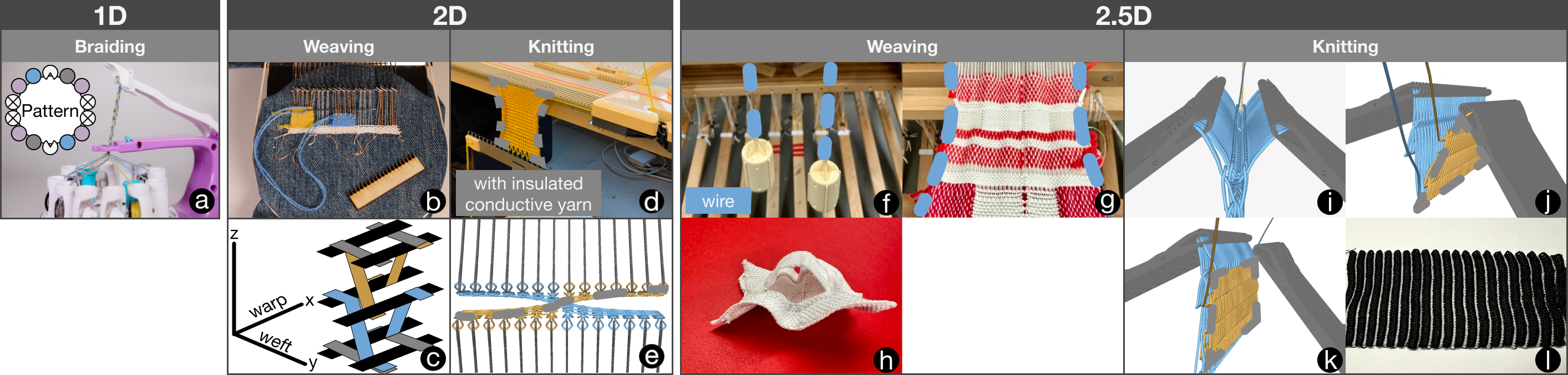}
  \caption{Techniques for fabricating (a) 1D yarn-based TENGs with a DIY braiding tool, (b) 2D single-layer woven structures, (c) 2D multilayer woven structures, (d) 2D single-layer knit structures, (e) 2D knit CS structures, (f-h) 2.5D woven structures on the floor loom, and (i-l) 2.5D Ottoman Stitch knit structures (blue, yellow, purple, and gray represent positive, negative, non-functional, and insulated conductive yarns, respectively.} 
  \label{fig:fabMethods}
  \Description{}
\end{figure*}

\subsection{Textile Structure: 2D TENG Structures (Figure~\ref{fig:DesignSpace}c)}
Weaving and knitting are emphasized within this design menu due to their capacity for precise control at the stitch level, allowing for pattern adjustments to enhance energy harvesting and the familiarity of textile tools for woven and knit construction. 
Yarns with triboelectric properties and \textit{SE yarns} are utilized in fabricating 2D textile structures.

Woven and knitted TENGs can be fabricated in \textit{SE} or \textit{CS mode} (Figure~\ref{fig:DesignSpace}c). In \textit{SE mode}, a tribo and conductive yarns are combined in a single-layer structure, or fabricated as two separate layers in a multilayer structure. In \textit{CS mode}, a multilayer structure, includes both positive and negative tribo and conductive yarns, are fabricated. In addition, two \textit{SE mode} with different polarities can be combined to create \textit{CS} or \textit{LS mode} TENGs.     

\subsubsection{Weaving}
Weaving offers a simple fabrication process, versatility with various yarns, and high integration performance. Commonly used weaves include plain, satin, and twill (Figure~\ref{fig:DesignSpace}c).
In our implementation, we utilized the plain weave for its resilience, abrasion resistance, and resistance to pilling~\cite{jahan2017effect}, and the 5/1 twill weave for its superior electrical performance~\cite{somkuwar2020structurally}, both in single-layer and multi-layer configurations.
We employed hand, darning, floor looms (8-shaft\footnote{\href{https://schachtspindle.com/products/standard-floor-loom-shaft-loom}{https://schachtspindle.com/products}}), and digital Jacquard loom (TC2\footnote{\url{https://digitalweaving.no/tc2-loom/}}) to facilitate the weaving process.
Although we utilized a TC2 for rapid prototyping, all weaving patterns can also be achieved with more readily available looms.
For single-layer \textit{SE textile}, we employed insulated conductive yarn as warps and tribo-yarns such as acrylic yarns as wefts.
We chose hand and darning looms for these cases due to their ease of warp replacement (Figure~\ref{fig:fabMethods}b). 
For applications where \textit{SE yarns} were used (e.g., core-shell or braided structure), we utilized cotton as the warp.
    
Weaving multi-layer \textit{SE textile} proved more efficient due to the increased surface area.
One of the simplest forms of this structure is the double weave construction (Figure~\ref{fig:DesignSpace}c)~\cite{Woven-eTextiles-in-HCI_Pouta_DIS22}.
For creating woven \textit{CS textile}, a three-layer structure with the tribo layers on the inside and two conductive layers on the outer sides would be suitable (Figure~\ref{fig:fabMethods}c and Supplementary S1). Alternatively, two \textit{SE textile} with different tribo polarities could be utilized for contact separation or lateral sliding applications (e.g., underarm application).

\subsubsection{Knitting}
We created knitted fabrics in both single and double layers using manual (Silver Reed SK280, LK150) and computerized (Shima Seiki SRY123) knitting machines. While we used a computerized knitting machine for efficiency, the suggested structures can also be fabricated with a manual knitting machine.
    
For single-layer \textit{SE textile}, we selected a jersey structure, combining tribo yarn and insulated conductive yarn in the same feeder (Figure~\ref{fig:fabMethods}d). For double-layer \textit{SE textile}, we created a tubular structure using a double system (two feeders knitting simultaneously) with a 16-gauge weft knitting bed on a computerized machine. This structure consists of two layers: one made from tribo yarn and the other from conductive yarn~\cite{underwood2009design}.

For \textit{CS textile}, we used a double system to create an alternating tubular structure, with each tube having a width of 1 cm. This configuration involves interlacing two layers of fabric together, alternating at regular intervals: one layer from tribo-positive yarn (blue), the other from tribo-negative yarn (yellow) with the insulated conductive yarn (gray) (Figure~\ref{fig:fabMethods}e and Supplementary S2)~\cite{DongKnittedTENGs}. It is possible to use two knitted \textit{SE textiles} with different polarities for contact separation or sliding mode applications.
    
\subsection{Textile Structure: 2.5D structures (Figure~\ref{fig:DesignSpace}d)}
    While 2D textile-based TENGs are relatively simple to fabricate, 2.5D structures can produce higher output power than 2D structures in close contact.
    This improvement is achieved by increasing the number of layers, incorporating both positive and negative tribo and conductive layers, and optimizing the spacing between them. Such modifications enhance energy harvesting, especially in \textit{CS mode}, and enable stretch/flex sensing functionalities. 
    Structural-level integration methods, such as weaving and knitting, facilitate the fabrication of 2.5D structures, requiring manual or computerized weaving and knitting machines. 

    \subsubsection{Weaving}
    For weaving 2.5D structures, we utilized the double weave~\cite{Woven-eTextiles-in-HCI_Pouta_DIS22} construction, employing a computer-controlled loom (Figure~\ref{fig:DesignSpace}d). The top layer consists of tribo yarn (e.g., PTFE (Lifetime, 107131)), while the bottom layer integrates an elastic tribo yarn (e.g., spandex (Hyosung, Creora H300)). To avoid pulling the edge of the swatch, we thread two wires through the comb on both sides of the swatch and suspend the wires from the front and back beams with weights (Figure~\ref{fig:fabMethods}f). The elastic yarn is then woven under tension with 25\% stretch (Figure~\ref{fig:fabMethods}g). After removing the swatch, the tension on the elastic yarn is released, transforming the textile into a 2.5D structure (Figure~\ref{fig:fabMethods}h and Supplementary S3). This created distance between layers enhances performance in \textit{CS mode}. 
    
    \subsubsection{Knitting}
    We knitted the 2.5D structure using the Ottoman stitch~\cite{ottoman-stitch} on a computerized knitting machine (Figure~\ref{fig:DesignSpace}-2.5D and Supplementary S4).
    The ottoman stitch, which can also be done on manual machines, forms a widthwise ribbed fabric with stitches between each rib.
    Initially, we knitted all needles on both beds using one elastic tribo yarn (e.g., sting (Silk City Fibers, 9500-S011)) as shown in Figure~\ref{fig:fabMethods}i. After a few courses, we continued knitting on the front bed using a second tribo yarn (e.g., PTFE) twisted with a conductive yarn, while holding the back stitches (Figure~\ref{fig:fabMethods}j). The number of courses of the front stitches determines the height of the 2.5D structure. Subsequently, we merged all needles by transferring the front stitches (Figure~\ref{fig:fabMethods}k) and repeated the knitting pattern. The released tension from the elastic yarn results in a 2.5D structure after removing the swatch from the beds (Figure~\ref{fig:fabMethods}l).

\subsection{Design Recommendation}

When selecting yarns for wearable devices, performance, sustainability, comfort, and compatibility with fabrication machines should be considered. Here are some recommendations:

\begin{itemize}
    \item \textbf{Core-Shell PTFE (-):} Highest performance, durability, UV resistance.
    \item \textbf{Lifetime PTFE (-):} Flexibility, durability, UV resistance.
    \item \textbf{Multifilament PTFE (-):} Softer, mimics human hair, ideal for skin contact applications.
    \item \textbf{Acrylic (-):} Sustainable, washable, recyclable alternative to PTFE.
    \item \textbf{Spandex (+):} Highest performance, preferred for its stretch and recovery properties, suitable for applications requiring elasticity and dimensionality.
    \item \textbf{Nylon (+):} Durability, flexibility, quick-drying.
    \item \textbf{Polyurethane (PU) (+):} Flexible and durable, ideal for high-tension fabrication processes.
    \item \textbf{Silk (+):} Soft, natural, biodegradable, and breathable, ideal for direct skin contact applications.
\end{itemize}

In addition, Yarn-based TENGs, particularly those created through braiding and twisting, are well-suited and visually appealing for accessories like hair extensions or bracelets. 
Knitted structures are preferable for applications requiring stretchability and flexibility, such as on joints or in socks. 
On the other hand, woven textiles provide stable and structured fabric, making them better suited for applications prone to abrasion and friction, requiring durability, such as body parts that rub against each other like the underarms and thighs. 
While 2D structures are appropriate for most applications, 2.5D structures excel in stretch/flex sensors, offering superior performance for contact separation mode due to the increased distance between tribo layers.

Given that \name{} was developed for the body, the best performance is achieved from body movements with higher frequency, force, contact area, and optimal distance between tribo-layers \cite{Zou_2020_Wearable}. In such scenarios, integrated TENG devices in socks or shoes (due to force and frequency) and positioned between the thighs or under the arms (due to the frequency and greater distance between two tribo-layers) yield the best results.
The forearm's smooth surface and accessibility presents an ideal location for touch-sensing applications enabling subtle interactions with the sensor. Stretch sensors, on the other hand, are most effective when placed at joints such as the elbow and knee, where significant movement occurs, or when monitoring muscle activity, such as in the case of human respiration. Additionally, strain gauges integrated into socks or shoe insole can serve as pedometers, counting the number of steps taken.
Once textile-based TENGs are fabricated, they can be connected to electronics through various methods, including snap fasteners \cite{SmartFabric_Castano_2014}.

\section{texTENG Applications} \label{application}

To demonstrate the practical feasibility of our approach, we present eight functional application examples (Figure~\ref{fig:teaser}). These exemplify diverse energy harvesting and actuation methods and a variety of self-powered sensors integrated into body locations using different materials and fabrication techniques. We also outline how each of these example builds on the design menu in Figure~\ref{fig:DesignSpace}.

\subsection{Energy Harvesting and Actuation}
Textile-based TENGs offer a seamless and continuous means of harvesting energy from human motion (e.g., walking, running, and muscle movements). The harvested energy is a sustainable power source for various applications, including powering small electronics, actuating devices, and charging batteries. The efficiency of energy generation depends on factors such as the size of the TENGs, materials used, and frequency of motion. Additionally, textile-based TENGs can be seamlessly integrated into everyday items like clothing and socks as demonstrated by the examples below.

\subsubsection{Powering wearable electronics: harnessing energy from the smart garment (Figure~\ref{fig:teaser}f).}

Utilizing the biomechanical energy from human motions such as walking or running, we crafted two swatches (\(100mm\times 100mm\)) and stitched them under the arm of a garment. As the arms swing, these swatches generate energy, directly and continuously powering a low-power wearable device like a digital watch (e.g., Casio, 0.3Wh), without the need for any battery.
The negative triboelectric swatch is composed of core-shell PTFE (Maedon, coated copper wire), woven in a single layer using a 5/1 twill structure on a computerized loom (TC2) (which can be fabricated on a floor loom as well). The positive triboelectric swatch is woven in a double-layer structure with nylon yarn (Tortoise, B0B7RFW927) on top and conductive yarn (GETREE, B0CNNKYP2C) on the bottom layer.
These swatches, functioning in contact-separation and sliding modes, are connected to the PMS circuit to power the digital watch.

\subsubsection{Illuminating running lights: harnessing energy with smart socks (Figure~\ref{fig:teaser}g).}
     To demonstrate the actuation capabilities of our approach, we created a pair of energy-harvesting socks inspired by \citet{DongKnittedTENGs}. Each sock generates an average peak voltage of \(13.2V\) and current of \(0.21 \mu A\) during running. The harvested energy is stored in a capacitor ($2.2\mu F$), which periodically discharges in short bursts, briefly flashing 20 series-connected low-power LEDs (Digikey, XZCM2CRK53WA-8VF) integrated into an armband (as shown in the supplementary video).   
    The socks' energy-harvesting layer (\(80mm\times 280mm\)) was crafted using a tubular knitting structure with PTFE yarn (Sailrite, 107131) twisted with conductive yarns (GETREE, B0CNNKYP2C) for the negative side and nylon yarn (Tortoise, B0B7RFW927) for the positive side, enabling effective energy generation through vertical contact separation mode. 
    Fabrication was done on a computerized knitting machine (Shima Seiki SRY123) or can be done on a manual knitting machine.
    The TENG layer is connected to a flexible PCB with a full-bridge rectifier (DB157) and LEDs using thin, flexible copper wire, yielding results comparable to those reported in \citet{DongKnittedTENGs}.

\subsection{Self-Powered Sensing}

Textile-based TENGs can function as self-powered wearable sensing systems, enabling continuous health monitoring and motion tracking without reliance on external power sources. By capturing signals generated from diverse mechanical movements, such as touching, running, and muscle contractions, various functionalities are achievable. Here we present six application examples of self-powered sensors.

\subsubsection{Subtle interaction with an interactive bracelet (Figure~\ref{fig:teaser}a).}

We developed an interactive bracelet designed for subtle interactions, such as sending emergency messages by simply pulling the elastic band. The fabrication process involved braiding silk (Woolery, FA-246) and acrylic (Mackellar) yarns around conductive yarns separately, creating two single \textit{SE yarns}. These yarns were then coiled around an elastic yarn (spandex (Hyosung, Creora H300)) to construct a \textit{CS yarn}. The fabricated sensor (\(80mm\)) is connected to the sensing board and communicates with a mobile phone via Bluetooth. The sensor reading for this application averages around \(1.7V\).

\subsubsection{Interactive hair extension (Figure~\ref{fig:teaser}b).}
To showcase the versatility of \name{}, we developed an interactive hair extension that can be used to activate voice assistants or trigger navigation commands when pressed.
The fabrication process involved twisting multifilament PTFE (Toray, 1200 denier) around a conductive yarn to create an \textit{SE yarn}. Subsequently, we braided the twisted PTFE yarn with core-shell PU (Maeden, coated copper wire) to form an interactive \textit{CS yarn} (\(200mm\)). The sensor reading for this application averages around \(1.8V\) when pressed.

\subsubsection{Smart home control with interactive drawstring (Figure~\ref{fig:teaser}c).}
We fabricated interactive touch-sensitive drawstrings that can be used for controlling smart home devices, such as RGB lights. We used the right drawstring to toggle the light on and off, while the left one adjusts the color, providing convenient control for users situated away from traditional switches.
These drawstrings are crafted by coating PDMS (Sylgard 186; Dow Corning (US)) onto a conductive thread, which is then braided with colored cotton for enhanced aesthetics. The resulting sensor, measuring \(20mm\), operates in single-electrode mode and detects finger touch, yielding an average output of \(3.1V\).

\subsubsection{Self-powered music control: seamlessly integrated touch sensors) (Figure~\ref{fig:teaser}e).}
    To demonstrate the seamless integration of \name{} devices into existing garments, we wove three self-powered touch sensors onto a coat sleeve (\(20mm\times 80mm\)).
    These sensors operate in single-electrode modes, functioning as music control buttons (Figure~\ref{fig:teaser}d).
    Using a darning loom with insulated conductive yarn (warp) and acrylic as the weft (Mackellar), we wove touch sensors connected to a sensing board in the user's pocket, which communicates with a music player via Bluetooth.
    The average output voltage for this application is around \(3.2V\).

\subsubsection{Game controller using a multi-dimensional flex sensor (Figure~\ref{fig:teaser}d)}
    To explore diverse structures within our design menu, we fabricated a 2.5D double-weave structure (\(100mm\times 30mm\)) using a floor loom. The sensor affixed to the wrist, functions as a flex sensor for game control (e.g., Flappy Bird). When the wrist flexes, the two layers make contact, generating signals. We employed core-shell PTFE yarn for the top layer and spandex yarn for the bottom layer. The sensor, linked to the sensing board, generated approximately \(1.18 V\) upon interaction.

\subsubsection{Wireless interaction using an elbow patch (Figure~\ref{fig:teaser}h)}

To demonstrate wireless interaction with \name{} devices, we knitted a 2.5D swatch with an Ottoman stitch pattern (\(100mm\times 50mm\)) on a computerized knitting machine (Shima Seiki) (which can be fabricated on a manual knitting machine). This sensor, designed as an elbow patch, was stitched onto a cloth. An electrode attached to a sofa armrest was connected to the sensing board. When the user places their hand on the armrest, the sensing circuit detects the generated signal through near-field electrostatic induction. This detected signal can be used to control room temperature and lighting based on the user's presence. We used PTFE lifetime yarns with Sting yarn, a nylon-spandex composite, to maximize the performance while maintaining elasticity suitable for knitting machines.
The average sensor reading for this application was \(0.95 V\).

\section{Hardware for texTENG}
\name{} utilizes two circuit designs to enable sensing and energy harvesting from the fabricated devices (Figure~\ref{fig:Hardware}).

\subsection{Hardware for Sensing}
\begin{figure}[h]
  \centering
  \includegraphics[width=\linewidth]{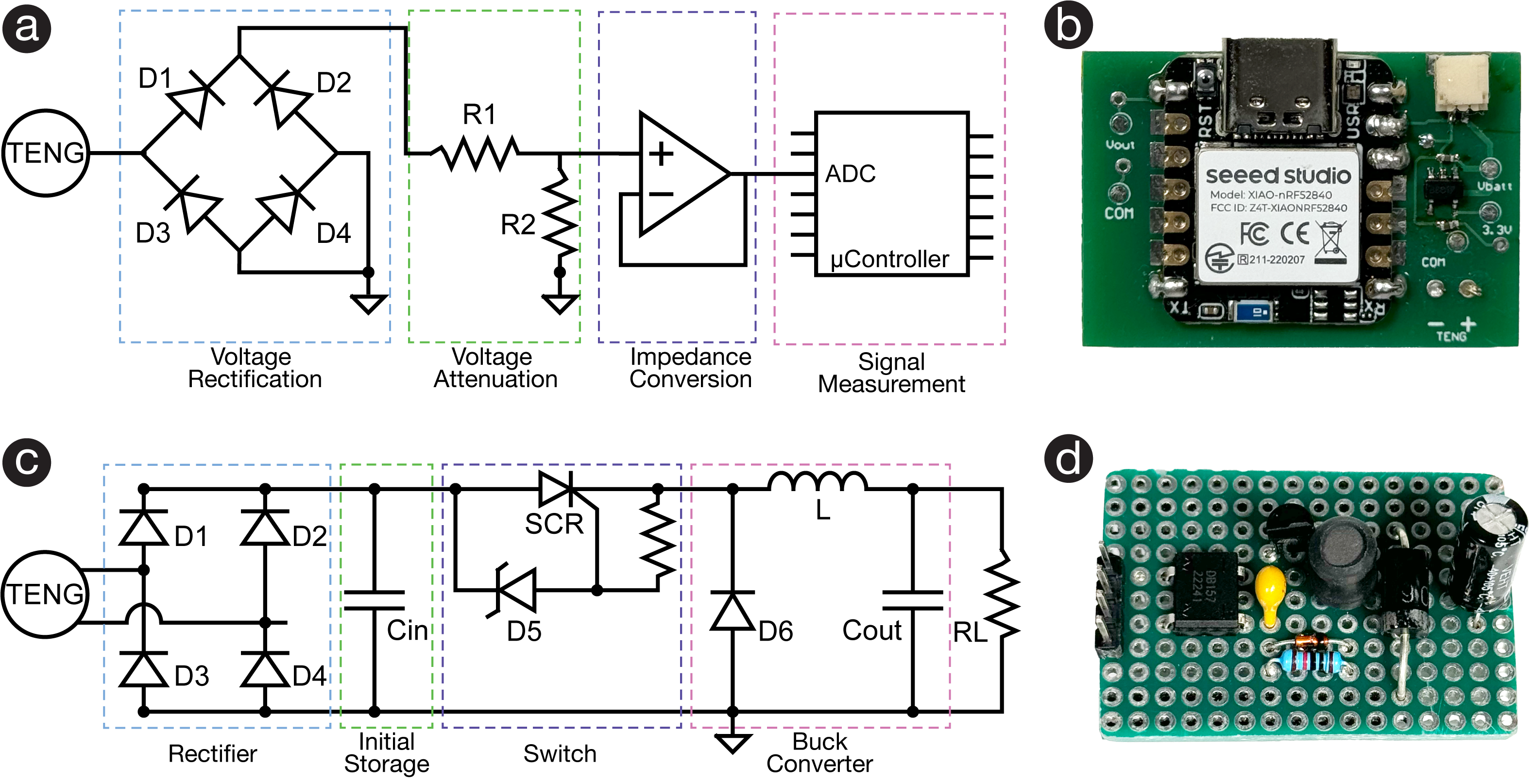}
  \caption{Circuit diagrams and PCB of our custom sensing board are shown in (a) and (b) and for the power management system (PMS) in (c) and (d). }
  \label{fig:Hardware}
  \Description{Hardware setup}
\end{figure}

    We designed a custom board (\(23mm\times 34mm\)) to measure the harvested signals from \name{}'s devices ~\cite{Skinergy_UIST23, lu2021regulating}. The board operates on an XIAO nRF52840 microcontroller (Seed Studio), equipped with Bluetooth Low Energy (BLE), and is powered by a $3.7V$ lipo battery (Figure~\ref{fig:Hardware} a,b).
    The sensing board consumes 37mW, but the fabricated sensors are self-powered and do not consume additional energy.
    The circuit consists of: 
    
    \noindent 
    \textbf{Voltage rectification:} full-bridge rectifiers (CDMBL102S) convert the harvested AC signals to DC signals (Figure~\ref{fig:Hardware}a). 

    \noindent 
    \textbf{Voltage attenuation}: incorporates a passive voltage divider network (R1 and R2) to reduce the signal magnitude to an acceptable range for the analog-to-digital converter (ADC) pins of the microcontroller. The values of the resistors are selected based on the input and desired output voltages (Figure~\ref{fig:Hardware}a). 
    
    \noindent 
    \textbf{Impedance conversion}: utilizes a rail-to-rail op-amp (MCP6V81) to convert high-impedance input signals to low-impedance signals, matching the impedance of the ADC pins (Figure~\ref{fig:Hardware}a). 
    
    \noindent 
    \textbf{Signal measurement}: the attenuated signals are measured using the microcontroller's ADC pins (Figure~\ref{fig:Hardware}a).

\subsection{Hardware for Harvesting}
    The output signals of TENGs typically exhibit high voltage and low current, making them unsuitable for direct use as a power source. To store the harvested energy from TENGs or directly power electronic devices, a power management system (PMS) is essential. The PMS ensures continuous and steady electrical voltage to power electronic devices. Following the proposed Self-driven PMS diagram by Harmon et al.~\cite{HarmonPowerManagementSystem}, we designed a custom board (\(23mm\times 40mm\)) based on a buck converter that does not require any additional power supply (Figure~\ref{fig:Hardware} c,d).
    The circuit consists of:

    \noindent        
    \textbf{Rectifier}: acts as a connection between the TENG and the energy storage system, converting the harvested AC signals to DC. We used a full bridge rectifier (DB157) (Figure~\ref{fig:Hardware}c).

    \noindent    
    \textbf{Initial storage}: serves as the primary energy storage unit following the rectifier. The voltage of $C_{in}$ sets a minimum voltage rating for the switch and $D_6$ (Figure~\ref{fig:Hardware}c). 

    \noindent    
    \textbf{Switch}: consists of a silicon-controlled rectifier (SCR, P0102DA) and a Zener diode (NZX2V1B). The switch activates when the initial storage voltage is at its peak; upon releasing the harvested energy to the subsequent part, the switch deactivates. The breakdown voltage of the Zener diode should be marginally below the input peak voltage to ensure that it is triggered after each peak input (Figure~\ref{fig:Hardware}c).
    
    \noindent
    \textbf{Buck converter}: comprises a parallel diode (1N5406), a series inductor ($L$), and a parallel capacitor ($C_{out}$) connected between the switch and the external load. When the switch is activated, the accumulated energy is stored in the LC unit. Conversely, when the switch deactivates, the energy stored in the LC unit is released to the load resistor (Figure~\ref{fig:Hardware}c).    
    
    The optimized values for our design, considering an input peak voltage of \(15 V\) at a frequency of \(1Hz\), are as follows: $C_{in}$ \(= 15pF\), $C_{out}$ \(= 47\mu F\), and $L$ \(= 2.3 mH\).
    We selected $C_{in}$ to match the input capacitance of our source, ensuring maximum power transfer as suggested in~\cite{xia2019universal}. Decreasing $C_{out}$ allows for fast charging and discharging.
\section{Evaluations}

We conducted a series of evaluations to understand the characteristics and performance of our textile-based TENGs.

\begin{figure*}[ht]
  \centering
  \includegraphics[width=\linewidth]{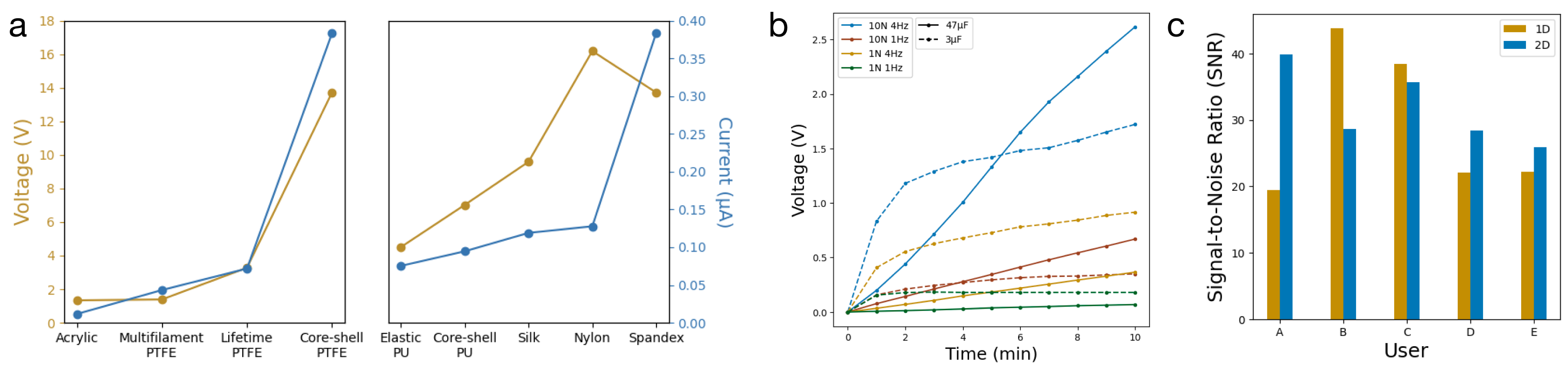}
  \caption{Evaluation results: a) material performance of nine commercially available yarns, b) assessment of energy storage capacity of PMS circuit, c) SNR ratio for touch input on 1D and 2D touch sensors. }
  \label{fig:evaluation-1}
  \Description{}
\end{figure*}

\subsection{Material Performance (Figure~\ref{fig:evaluation-1}a)}
    We evaluated the performance of various triboelectric materials within our design menu by comparing their open-circuit voltage ($V_{oc}$) and short-circuit current ($I_{sc}$) in contact-separation mode \cite{zou2019quantifying}. We fabricated nine woven swatches on hand looms (\(30mm\times 30mm\)) using commercially available triboelectric yarns, comprising of four tribo-negative: acrylic (Mackellar), multifilament PTFE (Toray, 1200 denier), lifetime PTFE (Sailrite, 107131), and core-shell PTFE (Maeden, coated copper wire) and five tribo-positive: elastic PU (Maeden, coated silver-copper alloy), core-shell PU (Maeden, coated copper wire), silk (Woolery, FA-246), nylon (Tortoise, B0B7RFW927), and spandex (Hyosung, Creora H300) yarns. Conductive stainless steel thread (Sparkfun) served as the warp, while tribo yarns were utilized as the weft. To prevent the shorting of electrodes upon contact with opposite polarity swatches, all tribo-positive swatches featured insulated conductive warps. For tribo yarns with conductive cores, such as those with a core-shell structure (e.g., core-shell PTFE), we employed cotton yarn as the warp due to its charge neutrality. 
    
    Drawing from literature~\cite{TriboelectricSeriesPUPTFE} and our own experiments, we identified swatches fabricated with core-shell PTFE and spandex as comparators for assessing the relative charge affinities of positive and negative swatches. To simulate the tapping effect in contact-separation mode, we selected a solenoid (12VDC, ADD06190484) with an impact force of \(1N\), replicating the force applied by users when interacting with smart devices \cite{NamForceTouchMeasurement}. The positive comparator was affixed to a 3D-printed extension of a push-pull solenoid, and the negative swatches were attached to a flat surface with \(5mm\) distance from each other. This process was repeated for all other positive swatches using our negative comparator.
    
    Performance metrics were recorded by capturing post-rectification $V_{oc}$ and $I_{sc}$ signals using an oscilloscope (Keysight DSOX2022A) and a desktop digital multimeter (Keithley DMM6500), respectively. 
    Peak voltage and current data were collected over 20 cycles at \(1Hz\) for each swatch pair, and averages were calculated.
    
    \noindent \textbf{Takeaway.} Figure~\ref{fig:evaluation-1}a illustrates that core-shell PTFE and spandex pairs exhibited the highest overall power performance, while acrylic and elastic PU pairs demonstrated lower performance.
 
\subsection{Energy Storage Capacity (Figure~\ref{fig:evaluation-1}b)}
    To assess the performance of the PMS circuit in storing energy, we conducted an evaluation. We selected two woven swatches (\(30mm\times 30mm\)): one fabricated with core-shell PTFE and the other with nylon as the weft and conductive stainless steel thread (Sparkfun) as the warp.
    One swatch was attached to the 3D-printed extension of a push-pull solenoid, while the other was positioned \(5mm\) away on a flat surface to assess the swatches in contact-separation mode.
    We subjected the evaluation to forces of \(1N\) (Solenoid, 12VDC, ADD06190484) and \(10N\) (Solenoid, 12VDC, A18051900UX0042) at two different frequencies (\(1Hz\) and \(4Hz\)), mimicking forces and frequencies typical of different body movements.
    Additionally, we varied the output capacitors (\(3\mu F\) and \(47\mu F\)) in the PMS circuit to observe their impact on the charging time. The swatches were connected to the PMS circuit, and the output voltage was measured every 30 seconds using a digital multimeter (FLUKE-115).   
    
    The results for 10 minutes are depicted in Figure~\ref{fig:evaluation-1}b. The evaluation revealed that increasing force and frequency led to faster and higher stored voltage, while increasing the output capacitor resulted in a slower charging time but also a higher voltage output and slower discharge. The best performance was achieved with a \(10N\) force, \(4Hz\) frequency, and a \(47\mu F\) output capacitor, reaching around \(1.7V\) after approximately 5 minutes.

   \noindent \textbf{Takeaway.} The results (Figure~\ref{fig:evaluation-1}b) show that high force and high-frequency motion resulted in faster and greater voltage storage. 
   This evaluation highlights the importance of selecting body locations for power harvesting that experience frequent and vigorous motion (e.g., underarm while running).

\subsection{Touch Sensitivity (Figure~\ref{fig:evaluation-1}c)}
    We conducted an evaluation to assess the robustness of the fabricated 1D and 2D touch sensors with input from different users. The 1D sensor utilized our PDMS-coated conductive yarn (length: \(100mm\)), while the 2D sensor was a hand-woven textile (size:\(30mm \times 30mm\)), fabricated with acrylic as the warp and conductive yarn as the weft. These sensors are connected to the sensing board and communicate with a laptop via Bluetooth.
    
    A controlled experiment involving five users (2 female) was conducted to formally investigate the signal-to-noise (SNR) ratio of the sensors and to account for variations in skin characteristics, such as hydration, known to vary across users. Participants were instructed to touch the 1D sensor with their index and thumb fingers and the 2D sensor with their index finger consecutively five times. The measured ADC values from our sensing board were recorded, and SNR values were calculated~\cite{Davison:2010:techniques} to gauge sensor performance.

       \noindent \textbf{Takeaway.}  The results (Figure~\ref{fig:evaluation-1}c) show that the SNR values for all participants are above 15, meeting the required threshold for robust touch sensing~\cite{Davison:2010:techniques}.
\section{Discussion, Limitations, and Future Work}

\noindent 
\textbf{Towards fully self-powered wearable devices.} 
In this paper, we investigated TENG wearable devices for self-powered sensing, energy harvesting, and actuation. 
In future work, exploring the use of supercapacitors to store harvested energy for later use could be beneficial~\cite{MAO2021105918}.
For \name{}, batteries powered the microcontroller for the sensing circuit, but aggregating energy from multiple devices and body motions could enable completely battery-free operation.
To improve performance, increasing swatch size, optimizing the PMS circuit~\cite{ZARGARI2021114489}, and incorporating techniques such as surface engineering~\cite{nanoenergyadv1010004} and electrospinning~\cite{bairagi2022high} offer paths to more efficient wearable energy harvesting systems.

\noindent 
\textbf{Wearability considerations for textile-based wearable TENGs.} 
Textile-based TENGs integrated into garments must prioritize wearability and comfort~\cite{TianTextileBasedTENGs}. 
This includes stretchability~\cite{DongFiberFabricTENG}, breathability~\cite{su2021self}, washability~\cite{feng2022highly}, and considerations of aesthetic appeal~\cite{younes2023smart} and social perception~\cite{gemperle1998designforwearability}.
For example, when selecting materials, PTFE-based fabrics demonstrate good resistance to water, making them suitable for textile-based TENGs that require frequent washing~\cite{NingWashable2018}.
However, there are tradeoffs between performance, ease of fabrication, and material cost that must be considered in material selection as introduced in our design space.
Future evaluations for long-term use of textile-based wearable TENGs could involve mechanical stretching tests (flexibility), air permeability tests (breathability), and repeated washing cycles (washability), as well as examining how makers choose materials based on the intended use cases.
Additionally, qualitative studies may examine the overall comfort, aesthetic, and user acceptance of textile-based TENGs and investigate how users may adapt to wearing devices that harvest energy from motion.

\noindent 
\textbf{Design Tools to support the making of textile-based power generators.} 
While this paper focuses on the fabrication and physical making of textile-based TENGs, the development of front-end software design tools for power and body location motion simulation could further benefit end users. The design tools could suggest suitable size, material, structure, and fabrication techniques, paired with body locations for generating the required amounts of power for various applications, which will encourage more sustainable prototyping practices within the HCI and maker community.
\section{Conclusion}

This paper introduced \name{}, a textile-based framework for wearable computing, leveraging triboelectric nanogenerators (TENGs) to harvest energy from human motion. Through a comprehensive exploration of its design menu, including fabrication techniques and integration strategies, \name{} demonstrates the versatility and adaptability of textile-based TENGs. 
Our example applications illustrate the diverse possibilities enabled by \name{}, from self-powered sensing to energy harvesting, underscoring its potential to enhance wearable device functionality while promoting sustainable energy sources. Technical evaluations confirm the performance and feasibility of \name{} devices, offering insights into touch input detection and energy storage efficiency. These contributions advance the understanding of textile-based TENGs, inspiring future developments in wearable computing and fostering innovation and sustainable prototyping within the HCI and maker communities.

\bibliographystyle{ACM-Reference-Format}
\bibliography{References}

\end{document}